# Google, How Should I Vote? How Users Formulate Search Queries to Find Political Information on Search Engines

Victoria Vziatysheva[1, *], Mykola Makhortykh[1], Maryna Sydorova[1] , Vihang Jumle[1]

**Abstract:** Search engine results depend not only on the algorithms but also on how users interact with them. However, factors affecting the selection of a search query remain understudied. Using a representative survey of Swiss citizens before a round of federal popular votes, this study examines how users formulate search queries related to the retirement policies that were voted on in March 2024. Contrary to existing research, we find no direct evidence of selective exposure, or users' tendency to search for pro-attitudinal information, which we explain by the less polarizing search topics. However, we find that the sentiment of the query is partially aligned with the expected vote outcome. Our results also suggest that undecided and non-voters are more likely to search for nuanced information, such as consequences and interpretations of the policies. The perceived importance and effect of the issue, political views, and sociodemographics also affect query formulation.

[1] Institute of Communication and Media Studies, University of Bern, Bern, Switzerland
* Corresponding author: Victoria Vziatysheva. Email: victoria.vziatysheva@unibe.ch | https://orcid.org/0000-0002-3762-6758

## Introduction

There has been a long-standing concern about the potential of algorithm-driven platforms to sway attitudes and opinions by prioritizing certain information, sources, or viewpoints to their users (Bruns, 2019). Search engines, often serving as primary access points to other sources, are especially powerful in shaping individual information diets. By filtering and ranking information, they can highlight certain topics, thus leading to agenda-setting effects (Lee et al., 2016). The choice of a search engine alone can determine which sources a person is directed to since different platforms produce notably different search results (Makhortykh et al., 2020). The output of search engines is, however, not free of systematic bias: for example, it was found that search results could reinforce or even amplify existing societal biases by underrepresenting women and non-white people (Guilbeault et al., 2024; Rohrbach et al., 2024).

However, exposure to information through search engines and its potential effects on opinions and decision-making do not solely depend on the algorithmic curation of the output but also on the user interactions with the specific content. For example, Robertson et al. (2023) found that even though search engines present news that is relatively balanced in terms of partisanship, users still primarily engage with identity-congruent websites, emphasizing "the role of user choice, rather than algorithmic curation" (p. 347). This behavior can be attributed to the phenomenon of selective exposure, or the tendency "both consciously and unconsciously to seek out material that supports their [people's] existing attitudes and opinions" (Chandler & Munday, 2011). Selective exposure has been extensively studied in communication research and documented in various contexts, including the consumption of news (Garrett, 2009), political advertising (Schmuck et al., 2020), and misinformation (Guess et al., 2018).

With search engines, however, selective exposure may occur not only in the stage of information selection (i.e., website visits) but also in the process of information seeking. In particular, how users conduct their searches and formulate search queries also affects the output that they get. Although selective exposure has been studied more in the context of the user clicking behavior rather than the context of search query selection, there is still evidence suggesting that users tend to prefer search queries that align with their pre-existing attitudes (Ekström et al., 2024). However, depending on the search topic, other individual factors may also play a significant role, such as perceived issue importance or political attitudes (van Hoof et al., 2024). Thus, it is important to consider user-side factors to understand how algorithms influence opinions and decision-making.

This paper examines how topic-specific attitudes, political views, and sociodemographic factors influence the selection and formulation of search queries in the context of Swiss semi-direct democracy, where citizens can vote on federal issues up to four times a year. We conducted a representative survey prior to a round of popular votes in March 2024 related to two retirement policies. This study investigates how citizens' voting intentions, perceptions of the proposed policies' impact and importance, expectations regarding the voting outcomes, political leanings, and other individual factors influence the formulation of search queries, particularly regarding their sentiment and subtopic.

Importantly, this is one of the few studies that investigate the information-seeking behavior of search engine users outside the US and focuses on less polarizing topics than those that are typically used in selective exposure or confirmation bias studies, such as migration, gun control or abortions (e.g., Ekström et al., 2024; Knobloch-Westerwick et al., 2015b; Slechten et al., 2022; Westerwick et al., 2017). Instead, we show how users formulate their search queries in the context of more specific and practical political issues (i.e., potential changes in pension legislation) directly related to political behavior (i.e., voting). This way, we aim to better understand information-seeking behavior in the context of routine online searches on political topics.

## Literature Review

### *Personalization of Political Information by Search Engines*

For over a decade, research has tried to determine whether algorithmic curation of content can lead to the formation of filter bubbles (Pariser, 2011), which could occur if algorithms exposed users mainly to content aligning with their beliefs and, therefore, isolated them from other viewpoints. Even though this phenomenon has found little empirical confirmation (Bruns, 2019), algorithmic platforms still personalize their output, which can lead to systematic differences in results for different groups of users. Generally, research shows that personalization by search engines is not extremely high: according to Hannak et al. (2013), it causes an 11.7% difference in search results. However, the degree of personalization varies depending on such factors as geolocation, date, or being logged into a Google account (Hannak et al., 2013; Kliman-Silver et al., 2015; Robertson et al., 2018).

Notably, the composition of search results can also be influenced by the search query selection. For instance, according to Kliman-Silver et al. (2015), queries mentioning local establishments and services (e.g., “bank” or “hospital”) led to more personalized results than queries mentioning politicians or controversial issues (e.g., climate change or same-sex marriage). The language of the query is another crucial driver of personalization, as evidence suggests that search engines promote different representations of issues such as historical atrocities (Makhortykh et al., 2022) or migration (Urman et al., 2022) for the same queries in different languages. At the same time, research suggests that search engines have “a mainstreaming effect” on the results: in particular, studying partisan search behavior among US users, Trielli and Diakopoulos (2022) find that search engines tend to recommend a similar set of results, even if queries reflect different political ideologies.

Using the search data collected during the month after Donald Trump’s inauguration, Robertson et al. (2018) explored whether the political leaning of users may implicitly affect the search results. They found no evidence that being a Democrat or a Republican led to greater personalization. However, participants who expressed low-strength attitudes toward Trump (scoring 1 or -1 on a -5 to 5 scale) received more personalized results than others. Taken together, these findings highlight not only the importance of algorithms but also the potential role of user-side factors (i.e., language, query selection, political views) that might contribute to the changes in the search engine output.

### *Selective Exposure in Political Searches*

Another strand of research explores how users interact with search results. It is well-known that top-ranked results typically attract the most clicks and viewing attention (e.g., Haas & Unkel, 2017; Pan et al., 2007; Zumofen, 2023). However, user preferences can also be driven by individual factors, such as interests, information needs, or existing beliefs.

There is mounting evidence that users tend to engage more with attitude-consistent information, a phenomenon also known as selective exposure or confirmation bias. For instance, a series of experiments by Knobloch-Westerwick et al. (2015a), Knobloch-Westerwick et al. (2015b), Westerwick et al. (2017), which used mock search results pages as stimuli, found that users prefer and spend more time with attitude-consistent messages, demonstrating the presence of selective exposure in online information-seeking behavior. Studies also show the effect of attitude reinforcement: engaging with attitude-consistent messages strengthened users’ attitudes toward an issue, whereas engaging with attitude-discrepant messages weakened them (Knobloch-Westerwick, et al., 2015b).

An eye-tracking experiment examining user search behavior (Ekström et al., 2022) found that both left- and right-wing participants tend to select pro-attitudinal links when reviewing search result pages. However, regarding visual attention, this effect was observed only for right-wing participants. Similarly, Robertson et al. (2023) found that voters with strong support for Republicans were significantly more likely to click on right-leaning sources in Google Search results, although there was no significant difference for voters with strong support for Democrats. However, the study found evidence of selective exposure for both political identities when examining overall engagement with websites (not only accessed through search engines).

At the same time, studies looking at broader online news consumption show that the use of search engines reduces selective exposure (Cardenal et al., 2019) and contributes to more diverse news repertoires (Fletcher et al., 2023). This suggests that while users might engage with attitude-consistent search results more often or more actively than with attitude-discrepant ones, in general, search engines still may lead users to more diverse information diets.

### *Formulation of Search Queries*

Up to now, studies have indicated several factors potentially influencing how a person phrases a search query. Using a sample of Dutch internet users, van Hoof et al. (2024) investigated the relationship between political attitudes and the formulation of queries related to immigration and climate change. Based on the search queries suggested by respondents, they identified different types of searchers and examined how attitudes toward the topics influence these types. For example, the authors found that immigration attitudes and political views affect query formulation for immigration-related topics—in particular, users with more positive attitudes toward immigration are more likely to use queries with a humanitarian framing. In contrast, for climate change searches, the issue importance appeared to be a more significant factor in query selection (van Hoof et al., 2024).

Ekström et al. (2024) explore the occurrence of "self-imposed filter bubbles" through search engines, referring to the tendency of users to select queries that align with their own views. In an experiment with 54 participants, the researchers found that when given a selection of queries on polarizing topics such as immigration, abortion, and sex equality, users tended to choose queries that matched their political leanings (measured on a left-right scale). A similar tendency was observed in a study by Blassnig et al. (2023), which, like the present paper, investigated how Swiss citizens search for information about upcoming referenda. By analyzing referenda-related queries obtained through data donation, the authors identified differences between proponents and opponents of the COVID-19 law. Proponents typically searched for information about the benefits or arguments in favor of the law, while opponents more frequently looked for the disadvantages and counterarguments. However, according to the study, the overall number of relevant queries was low (65), as referenda-related searches were generally quite rare.

In a similar vein, Menchen-Trevino et al. (2023) demonstrated that political search topics can, to some extent, reflect user partisanship (e.g., Democrats searching more for topics related to race and Republicans—for immigration). However, explicitly partisan language in the queries (collected via data donation) was very rare and, even when it did occur, was not necessarily aligned with the user's own political leaning. In another study, participants were prompted to search for information on various socio-political statements (Slechten et al., 2022). It was found that higher agreement with a statement predicted the use of queries that would confirm this belief, suggesting that the attitudinal strength can also affect to what degree the queries are aligned with preexisting attitudes.

In general, existing research presents a rather fragmented picture of the user selection of search queries. While there is evidence of selective exposure in query formulation, it is not straightforward, which, however, can be partially explained by substantial differences in the study designs and methodologies. Furthermore, factors affecting the query construction might vary depending on the search topic, which requires further investigation—also beyond the polarizing issues typically used for selective exposure studies.

## Hypotheses

Relying on the evidence of selective exposure discussed above, as well as the well-documented phenomena of confirmation bias and cognitive dissonance (Festinger, 1957), we assume that users will prefer search queries aligning with their beliefs. Since the study investigates web search behavior in the context of popular votes in Switzerland, we use voting intention—in favor or against a popular initiative to be voted on—as an indicator of an attitude towards an initiative. Thus, we propose the following hypotheses:

*H1a: Proponents of the initiative are more likely to use queries with a positive sentiment.*

*H1b: Opponents of the initiative are more likely to use queries with a negative sentiment.*

*H1c: Undecided voters or non-voters are more likely to use queries with a mixed sentiment.*

Studies show that the expectations regarding the vote outcome (e.g., how positive or negative it will be) can affect the voting decision (Fisher & Renwick, 2018; Grynberg et al., 2020). Following this proposition, we hypothesize that the effect that voters expect an initiative to have on them, if accepted, will influence the sentiment of a query.

*H2a: The more positive the effect voters expect from the initiative, the more likely they are to use queries with a positive sentiment.*

*H2b: The more negative the effect voters expect from the initiative, the more likely they are to use queries with a negative sentiment.*

Next, building on the notion of bandwagon effect, or the idea that voters are more likely to vote for the candidate that they expect to win (Kiss & Simonovits, 2014), we also assume that the expected outcome of the votes will be related to the query sentiment. This leads us to the following hypotheses:

*H3a: Voters expecting the initiative to be accepted are more likely to use queries with a positive sentiment.*

*H3b: Voters expecting the initiative to be rejected are more likely to use queries with a negative sentiment.*

Finally, we assume that the perceived importance of an initiative is related to the likelihood of searching for more nuanced aspects of an initiative.

*H4a: The higher the perceived importance of the initiative, the more likely a person is to look for its consequences.*

*H4b: The higher the perceived importance of the initiative, the more likely a person is to look for its interpretations.*

## Methods

### *Survey Sample*

We carried out a representative survey of 1,100 Swiss citizens recruited online by the Swiss social and market research company Demoscope. The survey was conducted in three major national languages in Switzerland: German, French, and Italian. The data were collected in early January 2024—two months before the round of federal popular votes, which was held on March 3, 2024.

During these votes, citizens voted for two popular initiatives:

(1) Initiative for a 13th OASI pension payment (also called “Better living in old age”; further referred to as the “OASI initiative”). It proposed an increase of the pensions paid from the Old Age and Survivors’ Insurance (OASI) foundation by one month’s worth (The Federal Council, 2024a). As a result of the vote, the initiative was accepted (58.3% yes votes).

(2) Pensions initiative (also called “For a secure and sustainable old-age pension scheme”). It proposed to increase the retirement age for women and men to 66 by 2033 and subsequently link it to life expectancy (The Federal Council, 2024b). As a result of the vote, the initiative was rejected (25.2% yes votes).

Popular initiatives are direct democratic tools to propose amendments to the Constitution. To be put up for vote, they must be supported by at least 100,000 signatures. Voter turnout for these issues was around 58%, which is higher than the average (Bundesamt für Statistik, n.d.).

The questionnaire included a set of questions about political leaning, voting intentions, attitudes toward the initiatives, search engine use, and sociodemographic characteristics. Respondents were also asked to provide three search queries for each of the popular initiatives that they would consider using to find more information about the votes.

### *Manual Coding of Search Queries*

To categorize the collected search queries, we designed a codebook (available in the Supplementary Materials) with a set of variables related to the query relevance, language, topic, and sentiment. Two student assistants fluent in at least one of the Swiss national languages completed the manual coding. Cohen’s kappa was calculated to assess the agreement between two coders. The kappa value varied between 0.68 and 1 for the variables of interest, indicating substantial agreement (Landis & Koch, 1977).

Then, we removed the queries that were considered meaningless. These included random sequences of letters, queries clearly indicating a non-response (e.g., “no”), and queries composed in languages other than German, French, and Italian (except for queries where the language was impossible to conclusively identify, e.g., if it contained a name of a platform or organization). This resulted in 6,427 queries eligible for the analysis.

### *Measures*

#### Independent Variables.

***Initiative Importance.*** Respondents were asked how important each of the initiatives was for them personally (from 1 = “Not at all important” to 5 = “Extremely important”).

***Perceived Effect of an Initiative.*** Measured by the question, “If the majority votes in favor, what will be the effect of the following proposals on you and your family?” (adapted from Pew Research Center, 2018), with response options 1 = “Very negative,” 2 = “Somewhat negative,” 3 = “Not much of an effect,” 4 = “Somewhat positive,” 5= “Very positive.”

***Voting Intention.*** Respondents were asked how they would vote for each of the initiatives (“How would you vote for the proposals if you were to take part in the popular votes on 3 March?”). For the analysis, their responses were grouped into three categories: (1) voting for, (2) voting against, and (3) undecided/non-voting.

***Voting Expected Outcome.*** Measured by the question “How do you think the proposals are likely to be voted on by the Swiss citizens?”. The responses were grouped into three categories: (1) for, (2) against, (3) 50-50.

***Political Interest.*** Respondents indicated their political interest on a 5-point Likert scale from 1 = “Not at all interested” to 5 = “Extremely interested.

***Political Leaning.*** Respondents were asked to place themselves on an 11-point scale where 0 = “Right” and 10 = “Left”.

***Political Efficacy.*** Four items in total, measuring internal (“I have a good understanding of the important political issues facing our country” and “I consider myself well qualified to participate in politics”) and external (“People like me don’t have any say in what the government does” and “No matter whom I vote for, it won't make a difference”) political efficacy were adapted from Ardèvol-Abreu et al. (2020) and Gil de Zúñiga et al. (2017). The items for external efficacy were reversed for the analysis, and the composite variable was created.

***Propensity to Vote.*** Respondents were asked to indicate how likely they are to “ever vote” for the six most prominent political parties (adapted from van der Eijk et al., 2006): Swiss People’s Party (further SVP), Socialist Democratic Party of Switzerland (further SP), FDP. The Liberals (further FDP), The Centre, the Green Party (further GP), and the Green Liberal Party (further GLP). Responses were measured on a scale from 0 = “I will certainly never vote for this party” to 10 = “I will certainly vote for this party at some time.”

In addition to the variables above, we controlled for age, gender, education as well as self-reported frequency of search engine use.

**Dependent Variables.**

***Sentiment.*** The sentiment of the query was broadly operationalized as the presence of any indicator of a positive/negative framing of an initiative. It was coded as follows: (1) Positive: a query positively portrays an initiative (e.g., "why initiative X is good") or seeks arguments in its favor (e.g., "initiative X pro," "initiative X advantages"); (2) Negative: a query negatively portrays an initiative (e.g., "why initiative X is bad") or seeks arguments against it (e.g., "initiative X contra," "initiative X disadvantages"); (3) Mixed: a query contains both positive and negative characteristics or seeks arguments both for and against an initiative (e.g., "initiative X pro and contra"); (4) Neutral: a query does not express any sentiment (e.g., an initiative name without additional details).

***Consequences.*** The queries were coded based on whether they referred to the implementation process of an initiative or its consequences. Coded as a binary variable: 1 = "yes," 0 = "no". Typically, such queries inquired about the financing of an initiative, its practical implications for a person, the initiative's effects on the economy, and other related issues.

***Interpretations.*** Measured whether a query inquires about arguments for or against an initiative (e.g., "X pro and contra") or different opinions on it (e.g., "X position of the parties"). Coded as a binary variable: 1 = "yes," 0 = "no".

## Data Analysis

Respondents with missing data on sociodemographic variables or younger than 18 were excluded. To handle missing values for other independent variables, we use a multiple imputation approach, which is more effective in better mitigating potential bias due to missing data (van Ginkel et al., 2020). This approach assumes imputing a missing value multiple times (instead of removing incomplete cases or imputing the value only once), thus, creating multiple complete datasets (Rubin, 1987). Missing values, in this case, are replaced by plausible values drawn from the distribution (van Buuren, 2018). The regression analysis is then performed on all of the datasets, after which estimates calculated for each of them are pooled into one. Unlike other ways of handling missing data, multiple imputation allows "to reflect uncertainty about which values to impute" (Rubin, 1987, p. 16).

Multiple imputation was carried out using the "mice" package in R with the classification and regression trees (CART) method (van Buuren, 2018). In the standard practice, data is imputed 2-10 times, which, however, may not be enough for the replicable *SE* estimates (von Hippel, 2020). Using the package "howManyImputations" (Errickson, 2024) based on the procedure suggested by von Hippel (2020), we determined that the optimal number of imputations for our data is 13.

To analyze the relationship between the features of the query and other studied factors, we ran a series of logistic regressions (a multinomial for sentiment and two binomial for consequences and interpretations). In this stage, we analyzed both initiatives together and paired queries on a specific initiative with attitudes towards a respective initiative (the initiative, however, was added as an independent variable). Initially, the same set of predictors without interactions was used for each dependent variable. However, we also assumed that the perceived initiative importance could moderate the effect of age (given that both initiatives are related to retirement) and other initiative-related variables (in particular, voting intention and the expected outcome). We gradually added these interactions to all three models. Having interactions of initiative importance with age and with voting intention (but not other variables) significantly improved the fit for the dependent variable "consequences," according to the results of the likelihood ratio test ($\chi^2(3) = 3.97$, $p = 0.008$). For sentiment and interpretations, interactions based on the initiative importance did not improve the models; thus, we only included them in the final model for consequences.

## Results

### *Descriptives*

Table 1 summarises the descriptive statistics for the respondent-level variables. The survey data, to a certain degree, mirrors the outcomes of the popular votes: the majority of the respondents (57.1%) had reported that they planned to vote for the OASI initiative, which was indeed accepted. The voting intentions regarding the Pensions initiative were split more evenly (40.9% for and 37.1% against); however, we observe that the majority (46.5%) expected the initiative to be rejected, which was the result of the votes. Furthermore, respondents indicated that they anticipated a more positive effect from the OASI initiative than from the Pensions initiative.

**Table 1. Descriptive statistics for the independent variables.**

| Variable | Summary |
| --- | --- |
| Gender | Female — 567 (51.5%)<br>Male — 533 (48.5%) |
| Age | M = 49.4, SD = 16.9 |
| Education | Low — 24 (2.2%)<br>Medium — 598 (54.4%)<br>High — 478 (43.4%) |
| Voting intention | **OASI initiative:**<br>For — 628 (57.1%)<br>Against — 275 (25 %)<br>Undecided/Not voting — 197 (17.9%) |
| | **Pensions initiative:**<br>For — 450 (40.9%)<br>Against — 408 (37.1 %)<br>Undecided/Not voting — 242 (22%) |
| Voting expected outcome | **OASI initiative:**<br>For — 596 (54.2%)<br>Against — 192 (17.5 %)<br>50-50 — 217 (19.7%)<br>NA = 95 (8.6%) |
| | **Pensions initiative:**<br>For — 255 (23.2%)<br>Against — 511 (46.5 %)<br>50-50 — 223 (20.3%)<br>NA = 111 (10%) |
| Initiative importance (1-5) | **OASI initiative:** M = 4.3, SD = 1<br>NA = 23 |
| | **Pensions initiative:** M = 4.2, SD = 1.1<br>NA = 27 |
| Perceived effect of an initiative (1-5) | **OASI initiative:** M = 3.8, SD = 1.1<br>NA = 77 |
| | **Pensions initiative:** M = 3.0, SD = 1.1<br>NA = 79 |
| Political interest (1-5) | M = 3.7, SD = 1.1<br>NA = 3 |
| Political leaning (0-10) | M = 4.7, SD = 2.3<br>NA = 94 |
| Political efficacy (1-5) | M = 3.5, SD = 0.8<br>NA = 46 |
| Propensity to vote (0-10) | **The Centre:** M = 4.8, SD = 2.7<br>NA = 91 |
| | **FDP:** M = 4.4, SD = 2.9<br>NA = 91 |
| | **GLP:** M = 4.3, SD = 3.0<br>NA = 83 |
| | **GP:** M = 4.3, SD = 3.3<br>NA = 70 |
| | **SP:** M = 5.0, SD = 3.3<br>NA = 74 |
| | **SVP:** M = 3.3, SD = 3.5<br>NA = 63 |

Note: for the subsequent regression analysis, missing values were imputed through multiple imputations; initiative-specific variables were paired with corresponding queries and analyzed together.

Despite the data being representative of the Swiss population based on sociodemographic parameters, we observe a discrepancy between the reported propensity to vote for major political parties and the actual support of these parties among the population. Notably, respondents reported the lowest propensity to vote for the SVP (M = 3.3 on the 11-point scale)—a right-wing populist party with the largest share in the Swiss Federal Assembly (27.9%). In our survey, respondents expressed the most positive attitudes towards the Social Democratic Party (SP), which had the second-highest number of votes in the last election in 2023 (18.3% share in the Federal Assembly).

Table 2 presents the results of the manual coding of the queries. The considerable majority of them did not express any explicit sentiment and were coded as neutral (89.7%). This category typically included queries that only consisted of an initiative name or its parts as well as general reference to the popular votes. The second most frequent sentiment was "mixed" (3.9%), typically represented by queries asking about the positive and negative sides of an initiative. Thus, expressing any clear attitude towards an initiative in a query was a relatively rare event. 15.6% of the queries inquired about possible consequences and implications of an initiative, whereas 15.1% were focused on interpretations (e.g., opinions on an initiative or arguments pro and contra).

**Table 2. Descriptive statistics for the dependent variables.**

| Variable | Category | Count | Percentage |
|---|---|---|---|
| Sentiment | Neutral | 5765 | 89.7% |
| | Positive | 221 | 3.4% |
| | Negative | 191 | 3% |
| | Mixed | 250 | 3.9% |
| Consequences | Yes | 1003 | 15.6% |
| | No | 5424 | 84.4% |
| Interpretations | Yes | 973 | 15.1% |
| | No | 5454 | 84.9% |

***Query Sentiment***

To test hypotheses related to the sentiment of the query, the multinomial logistic regression was used. As the reference level of the dependent variable, we chose a neutral sentiment representing the largest group of the queries; thus, all other types of sentiment were compared against it. The results for all regressions are available in the Supplementary Materials.

According to our results (Fig 1), there is no significant relationship between voting intention and query sentiment. In other words, initiative opponents are not more likely to use negative queries, and proponents are not more likely to use positive ones. We also do not observe any significant difference in query sentiment for undecided and non-voters. Thus, hypotheses H1a, H1b, and H1c are rejected. The perceived effect of an initiative is also non-significant in this model, which contradicts H2a and H2b.

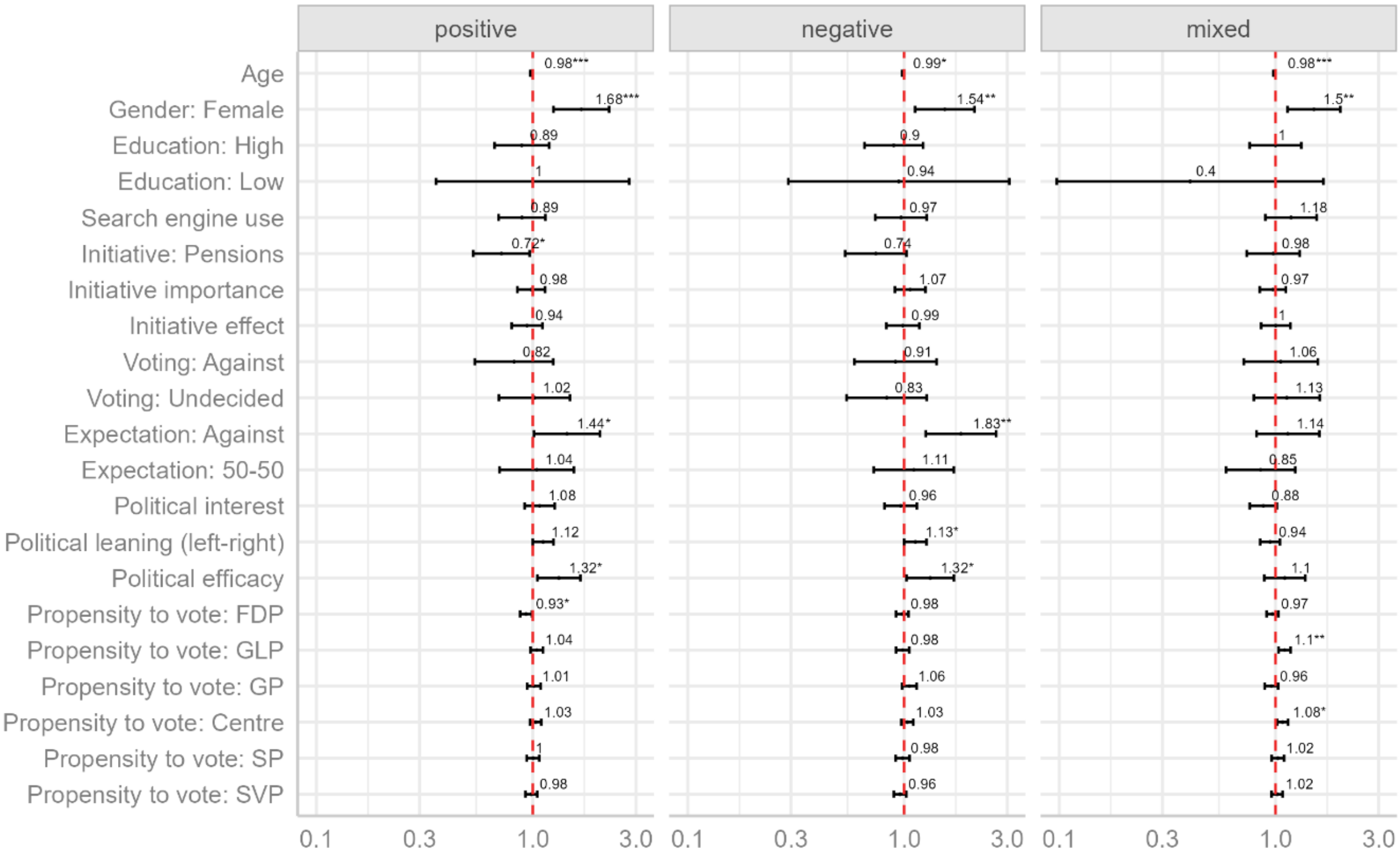

**Figure 1. Odds ratios of Sentiment across predictors.** *** $p < 0.001$, ** $p < 0.01$, * $p < 0.05$.

The hypothesis regarding the expected outcome of votes is supported partially. In particular, we found that respondents who expect an initiative to be rejected are more likely to use negative queries, with this relationship being rather strong (OR = 1.83, 95% CI: 1.26-2.66, $p < 0.01$) (H3b confirmed). However, we find the same effect for positive queries: they are more likely to be used by people expecting an initiative to be rejected and not by those who expect an initiative to be accepted (H3a rejected).

Age is negatively associated with using queries expressing any sentiment (Mixed: OR = 0.99, 95% CI: 0.98-0.99, $p = 0.001$; Positive: OR = 0.98, 95% CI: 0.97-0.99, $p < 0.001$; Negative: OR = 0.99, 95% CI: 0.98-0.99, $p < 0.001$), indicating that the older the respondent, the more likely they to use neutral queries. However, as the odds ratio is close to 1, this effect can be treated as relatively small, albeit significant. Similarly, we find that women are significantly more likely than men to use non-neutral queries (Mixed: OR = 1.51, 95% CI: 1.14-2, $p < 0.01$; Positive: OR = 1.68, 95% CI: 1.25-2.26, $p = 0.001$; Negative: OR = 1.54, 95% CI: 1.12-2.11, $p < 0.01$).

Looking at other predictors, we found that queries related to the Pensions initiative are less likely to express positive sentiment (OR = 0.72, 95% CI: 0.53-0.97, $p < 0.05$) than those related to the OASI initiative. Our results also show that the sentiment of the query is, to a certain degree, related to political attitudes. In particular, higher political efficacy provides a moderate explanation for the use of both positive and negative queries (Positive: OR = 1.32, 95% CI: 1.05-1.66, $p < 0.05$; Negative: OR = 1.32, 95% CI: 1.03-1.7, $p < 0.05$). Furthermore, the use of queries with negative sentiment is marginally associated with the more left positioning on the political leaning scale (Negative: OR = 1.13, 95% CI: 1.001-1.27, $p < 0.05$). Finally, expressing a mixed sentiment is positively associated with a propensity to vote for The Centre (OR = 1.08, 95% CI: 1.02-1.14, $p < 0.05$) and the GLP (OR = 1.1, 95% CI: 1.03-1.17, $p < 0.01$), and expressing a positive sentiment is negatively associated with a propensity to vote for FDP (OR = 0.93, 95% CI: 0.87-0.99, $p < 0.05$), however these effects are, again, rather small.

### *Consequences*

Next, we used binomial regression to examine the factors associated with searching for the consequences of an initiative. This model additionally included two interactions of initiative importance with age and voting intention, as they significantly improved the model’s fit for the consequence variable (but not for sentiment and interpretations).

First, we looked at the effect of the perceived importance of initiative. The results show (Fig 2) that respondents for whom an initiative is more important are indeed more likely to look for its consequences (OR = 1.47, 95% CI: 1.12-1.93, $p < 0.01$). Thus, H4a was confirmed.

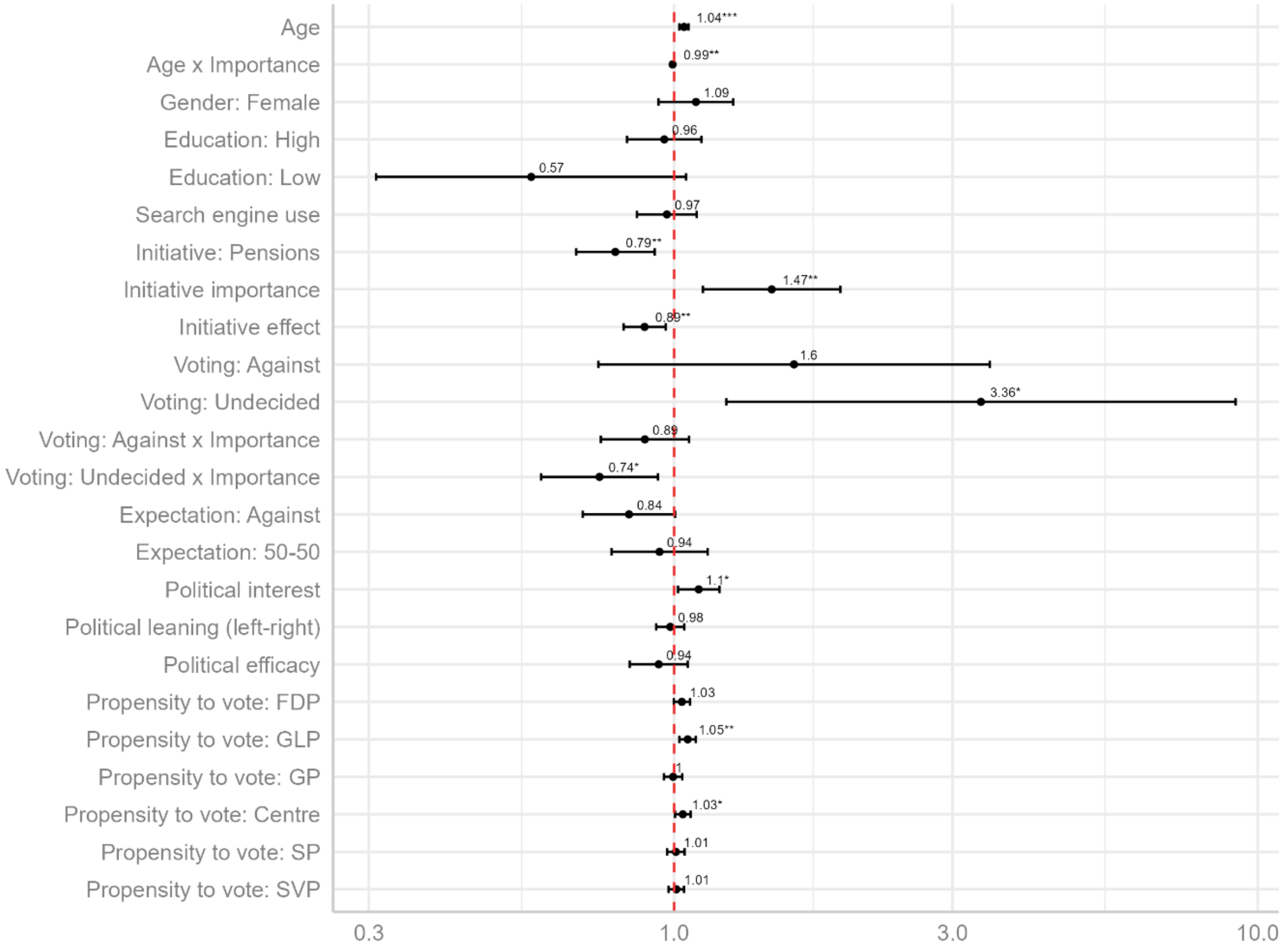

**Figure 2. Odds ratios of Consequences across predictors.** *** $p < 0.001$, ** $p < 0.01$, * $p < 0.05$.

Generally, respondents are less likely to inquire about the consequences of the Pensions initiative than about the consequences of the OASI initiative (OR = 0.79, 95% CI: 0.68-0.93, $p < 0.01$). This can be attributed to the fact that respondents mostly did not expect the Pensions initiative to pass and, thus, might have been less concerned about its implications.

Age appeared to be a positive predictor of searching for consequences (OR = 1.04, 95% CI: 1.02-1.06, $p < 0.001$); however, as with sentiment, the effect was small. We also observe that the interaction of age and importance has a different directionality from the one for these predictors separately. Specifically, older participants who perceive an initiative as more important are *less* likely to search for the consequences of this initiative (OR = 0.99, 95% CI: 0.99-0.998, $p < 0.01$), although this relationship is also relatively marginal.

Furthermore, we find that undecided and non-voters are considerably more likely to look for consequences (OR = 3.36, 95% CI: 1.23-9.17, $p < 0.05$) than those voting in favor of an initiative (the reference level). This can be explained by the fact that citizens who do not have a specific voting decision may require more context for an issue. This effect is, however, reversed by the perceived importance. That is, undecided and non-voters considering an initiative to be important are less likely to look for its consequences (OR = 0.75, 95% CI: 0.59-0.94, $p < 0.05$). This could be explained by these citizens being more informed about an initiative than the undecided voters in general, despite not planning or not knowing how to vote yet.

The analysis also shows that the more negatively the expected effect of an initiative is perceived, the more likely respondents to search for its consequences (OR = 0.9, 95% CI: 0.83-0.98, $p < 0.05$). Political interest was positively related to mentioning consequences in a query (OR = 1.1, 95% CI: 1.02-1.2, $p < 0.05$). Finally, the propensity to vote for The Centre (OR = 1.04, 95% CI: 1-1.07, $p < 0.05$) and GLP (OR = 1.06, 95% CI: 1.02-1.09, $p < 0.01$) is positively associated with searching for the consequences of an initiative.

***Interpretations***

Similarly to consequences, undecided and non-voters are more likely to look for interpretations of an initiative (OR = 1.31, 95% CI: 1.08-1.59, $p < 0.01$), which again can be attributed to the higher need for additional information (Fig 3). Here, we also observed the small effect of age, which is negatively associated with searching for interpretations of an initiative (OR = 0.99, 95% CI: 0.985-0.994, $p < 0.001$). Searching for interpretations is positively related to political efficacy (OR = 1.25, 95% CI: 1.11-1.4, $p < 0.001$) and negatively — to the propensity to vote for FDP (OR = 0.96, 95% CI: 0.93-0.99, $p < 0.05$).

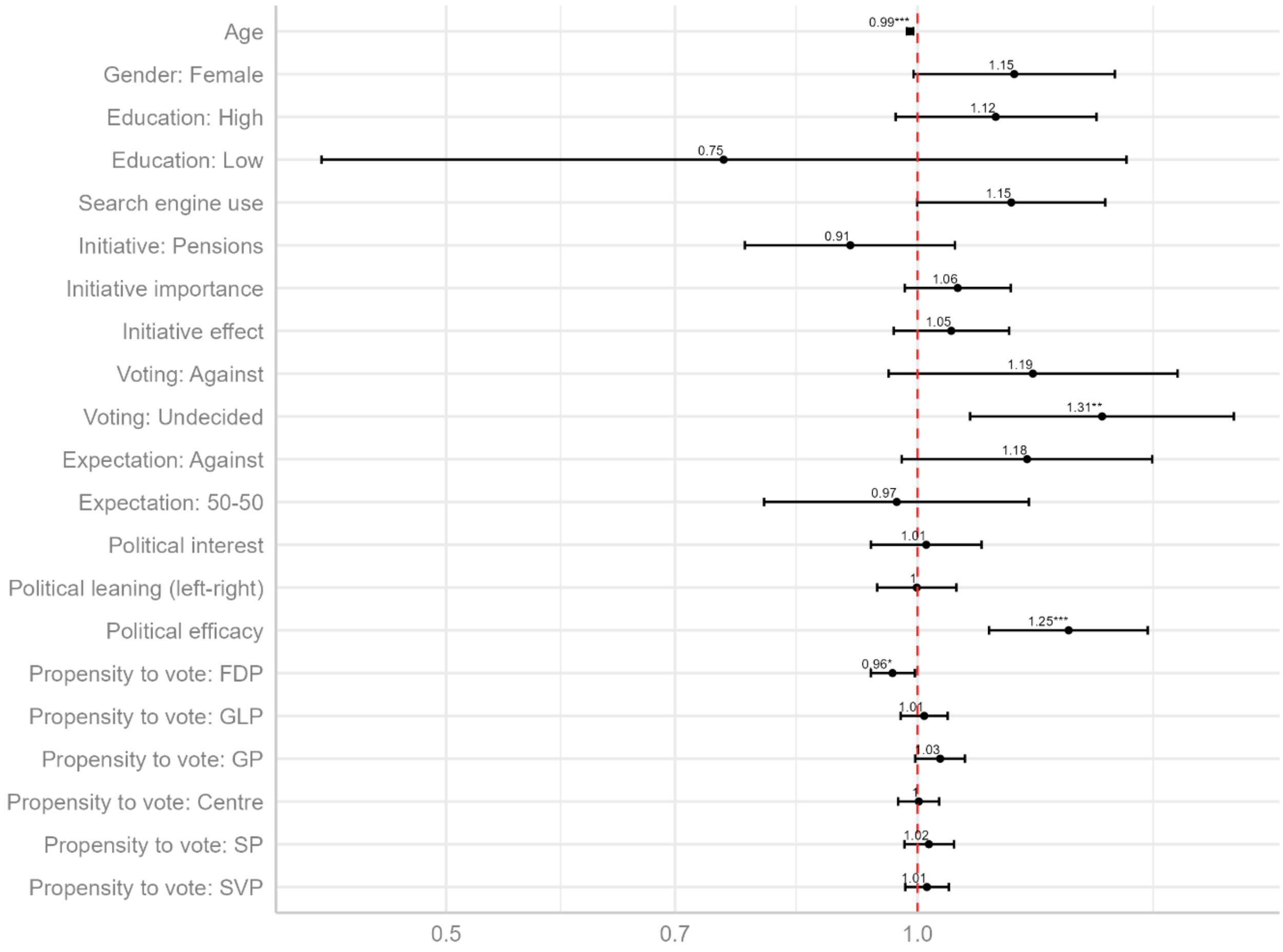

**Figure 3. Odds ratios of Interpretations across predictors.** *** p < 0.001, ** p < 0.01, * p < 0.05.

## Discussion

As algorithm-driven systems such as search engines offer more personalization and become increasingly sensitive to user input, it is critical to understand what influences user interactions with these platforms, particularly query formulation. This question has become especially important with the rise of generative AI and LLM-powered chatbots, which can serve as an alternative to search engines. However, it is known that due to their stochasticity, even slight changes in the prompt can lead to significant differences in results (Röttger et al., 2024). This highlights the growing importance of studying how users select and phrase their search requests.

This study aims to fill this gap by investigating how users formulate their search queries to find political information in the context of Swiss popular votes. It expands the understanding of individual factors affecting user selection of search queries, which, as we argue, remain relatively understudied, especially in contexts different from the US and for not severely polarizing political issues. The study yields several significant findings.

One of the main assumptions underlying this research was that users would choose queries reinforcing their beliefs, which leads to selective exposure. However, contrary to our hypotheses and existing evidence (e.g., Ekström et al., 2024), our findings do not indicate the signs of selective exposure in the formulation of search queries. In particular, proponents and opponents of the initiatives are not more likely to compose pro-attitudinal queries. Furthermore, the perceived effect of the initiative is not associated with the query sentiment; in other words, voters who expect an initiative to have a negative outcome are not more likely to select negative queries and vice versa. Thus, this study does not find clear evidence of the user's tendency to confirm their own views in the process of information-seeking behavior on search engines. However, this discrepancy with the existing research might be, to a certain degree, explained by the political issues chosen as a topic of search. While many studies on selective exposure focus on highly polarizing topics, we asked respondents to compose search queries for issues that are likely to elicit less polarization. This might indicate that selective exposure might not necessarily be present for the less controversial socio-political issues.

We, however, find that respondents expecting an initiative to be rejected tend to select negative queries more often. On the one hand, we can partially attribute it to a bandwagon effect: in other words, citizens who think that the majority does not support an initiative tend to adopt this belief and consequently portray it more negatively. However, within the scope of this study, it might be difficult to disentangle the bandwagon effect from the projection effects (i.e., individuals projecting their own beliefs on others) (Schmitt-Beck, 2015). Therefore, this evidence might also be attributed to selective exposure to a certain degree. Furthermore, this finding is inconclusive, as respondents expecting an initiative to be rejected are also more likely to use positive queries than those who expect an initiative to be accepted (opposite to our initial hypothesis).

Next, the study demonstrates that the composition of search queries depends on the issue-specific attitudes and corresponding information needs. For example, we find that searching for consequences is, among other things, explained by the higher perceived importance of an initiative. This result is aligned with the findings of van Hoof et al. (2024) on the searches related to climate change. Furthermore, undecided and non-voters are more likely to search for consequences and interpretations of an initiative, which indicates a need for more context than for those who have already made their voting decision. This finding is especially important, as search engines can potentially lead to a stronger attitude shift among undecided voters (Epstein & Robertson, 2015).

Sociodemographic characteristics also play a role in the query selection, with age being associated with all of the query features, albeit these effects are relatively small. We observe that the older the respondents, the more likely they are to look for the consequences of an initiative and the less likely they are to express any clear sentiment in their queries or look for interpretations. One of the possible explanations is the nature of the initiatives, both of which are related to retirement. Hence, we can assume that these proposals are more relevant for older people who might have had an opinion on them early on (and, thus, are less interested in interpretations) but for whom practical aspects of the initiatives are the most crucial (hence, higher interest in the consequences). We also find the effect of gender, with women being significantly more likely to select non-neutral queries than men. This, again, can be partly explained by the focus of the initiatives: currently, there is a considerable gap in the amount of the pension payment, with women receiving much less than men (Bundesamt für Statistik, 2024). Moreover, the increase in the retirement age for women (from 64 to 65, as it is now for men) has already been passed in 2022. This makes the impact of the initiatives potentially stronger for women than for men, leading to more queries expressing a certain attitude towards an initiative.

Finally, general political attitudes also partially explain the choice of queries. For instance, users with higher political efficacy are more likely to use both positive and negative queries, as well as look for the interpretations of an initiative. This might indicate that respondents believing they have more influence on political processes tend to select more nuanced queries and look for diverse political opinions. Searching for the consequences is, on the other hand, related to higher political interest. We also found that the propensity to vote for specific political parties is associated with certain query features. Although with a given approach to data analysis, we cannot trace the alignment of queries with the party ideology (as two initiatives were analyzed together and parties could have had different stances on them), it still indicated the importance of political attitudes for information-seeking behavior.

This study, however, has several limitations. First of all, the queries studied are self-reported and do not necessarily fully represent the actual searching behavior (e.g., with this design, we cannot take into account query suggestions and auto-completion by the search engines). Secondly, in line with the study by Blassnig et al. (2023), the majority of the collected queries are neutral and do not have much variability in terms of the searched subtopics and aspects of the initiatives. Thus, for instance, queries expressing sentiment other than neutral or looking for consequences and interpretations of an initiative comprise a smaller (even though still considerable) part of our sample. Finally, conducting our analysis on the level of an individual query, we do not account for the general composition of queries per respondent.

These findings indicate several directions for future research. Further investigation into the information-seeking behavior related to less polarizing political topics is needed. As we demonstrate in the present paper, the patterns of online search for more event-specific and developing topics might diverge from the search related to more general political issues, such as climate change. However, with Switzerland being a very particular case of a semi-direct democracy, our results should be treated carefully if generalized to other political and national contexts. Thus, we recommend a more nuanced selection of topics for studying information-seeking behavior in different national, regional, and political contexts (especially those in the Global South). Furthermore, our research highlights the importance of studying a more nuanced set of individual factors that affect the selection and formulation of search queries. In particular, it is crucial to capture factors specific to each given issue (e.g., pre-existing knowledge, relevance, etc.) and to go beyond a dichotomous understanding of pro- or contra-attitudinal information-seeking behavior.

## Data Availability

The data underlying this article will be shared on reasonable request to the corresponding author.

## Funding

This article is part of the project “Algorithm audit of the impact of user- and system-side factors on web search bias in the context of federal popular votes in Switzerland” (PI: Mykola Makhortykh) funded by the Swiss National Science Foundation [105217_215021].

## Ethical approval

The design of the project has been approved by the Ethics Committee of the Faculty of Business, Economics and Social Sciences of the University of Bern (serial number 382023).

## References

Ardèvol-Abreu, A., Gil de Zúñiga, H., & Gámez, E. (2020). The influence of conspiracy beliefs on conventional and unconventional forms of political participation: The mediating role of political efficacy. *British Journal of Social Psychology*, *59*(2), 549–569. https://doi.org/10.1111/bjso.12366

Blassnig, S., Mitova, E., Pfiffner, N., & Reiss, M. V. (2023). Googling referendum campaigns: Analyzing online search patterns regarding Swiss direct-democratic votes. *Media and Communication*, *11*(1), 19–30. https://doi.org/10.17645/mac.v11i1.6030

Bruns, A. (2019). *Are filter bubbles real?* Polity. https://www.wiley.com/en-us/Are+Filter+Bubbles+Real%3F-p-9781509536443

Bundesamt für Statistik. (n.d.). *Stimmbeteiligung*. Retrieved September 26, 2024, from https://www.bfs.admin.ch/bfs/de/home/statistiken/politik/abstimmungen/stimmbeteiligung.html

Bundesamt für Statistik. (2024). *Pension gap*. https://www.bfs.admin.ch/bfs/de/home/statistiken/wirtschaftliche-soziale-situation-bevoelkerung/gleichstellung-frau-mann/einkommen/pension-gap.html

Cardenal, A. S., Aguilar-Paredes, C., Galais, C., & Pérez-Montoro, M. (2019). Digital technologies and selective exposure: How choice and filter bubbles shape news media exposure. *The International Journal of Press/Politics*, *24*(4), 465–486. https://doi.org/10.1177/1940161219862988

Chandler, D., & Munday, R. (2011). Selective exposure. In *A Dictionary of Media and Communication*. Oxford University Press. https://www.oxfordreference.com/display/10.1093/acref/9780199568758.001.0001/acref-9780199568758-e-2402

Ekström, A. G., Madison, G., Olsson, E. J., & Tsapos, M. (2024). The search query filter bubble: Effect of user ideology on political leaning of search results through query selection. *Information, Communication & Society*, *27*(5), 878–894. https://doi.org/10.1080/1369118X.2023.2230242

Ekström, A. G., Niehorster, D. C., & Olsson, E. J. (2022). Self-imposed filter bubbles: Selective attention and exposure in online search. *Computers in Human Behavior Reports*, *7*, 100226. https://doi.org/10.1016/j.chbr.2022.100226

Epstein, R., & Robertson, R. E. (2015). The search engine manipulation effect (SEME) and its possible impact on the outcomes of elections. *Proceedings of the National Academy of Sciences*, *112*(33), E4512–E4521. https://doi.org/10.1073/pnas.1419828112

Errickson, J. (2024). *howManyImputations: Calculate How many imputations are needed for multiple imputation* (Version 0.2.5) [R package]. https://CRAN.R-project.org/package=howManyImputations

Festinger, L. (1957). *A theory of cognitive dissonance*. Redwood City: Stanford University Press. https://doi.org/10.1515/9781503620766

Fisher, S. D., & Renwick, A. (2018). The UK's referendum on EU membership of June 2016: How expectations of Brexit's impact affected the outcome. *Acta Politica*, *53*(4), 590–611. https://doi.org/10.1057/s41269-018-0111-3

Fletcher, R., Kalogeropoulos, A., & Nielsen, R. K. (2023). More diverse, more politically varied: How social media, search engines and aggregators shape news repertoires in the United Kingdom. *New Media & Society*, *25*(8), 2118–2139. https://doi.org/10.1177/14614448211027393

Garrett, R. K. (2009). Echo chambers online?: Politically motivated selective exposure among Internet news users1. *Journal of Computer-Mediated Communication*, *14*(2), 265–285. https://doi.org/10.1111/j.1083-6101.2009.01440.x

Gil de Zúñiga, H., Diehl, T., & Ardévol-Abreu, A. (2017). Internal, external, and government political efficacy: Effects on news use, discussion, and political participation. *Journal of Broadcasting & Electronic Media*, *61*(3), 574–596. https://doi.org/10.1080/08838151.2017.1344672

Grynberg, C., Walter, S., & Wasserfallen, F. (2020). Expectations, vote choice and opinion stability since the 2016 Brexit referendum. *European Union Politics*, *21*(2), 255–275. https://doi.org/10.1177/1465116519892328

Guess, A., Nyhan, B., & Reifler, J. (2018). Selective exposure to misinformation: Evidence from the consumption of fake news during the 2016 US presidential campaign. *European Research Council*, *9*(3), 4. https://about.fb.com/wp-content/uploads/2018/01/fake-news-2016.pdf

Guilbeault, D., Delecourt, S., Hull, T., Desikan, B. S., Chu, M., & Nadler, E. (2024). Online images amplify gender bias. *Nature*, 1–7. https://doi.org/10.1038/s41586-024-07068-x

Haas, A., & Unkel, J. (2017). Ranking versus reputation: Perception and effects of search result credibility. *Behaviour & Information Technology*, *36*(12), 1285–1298. https://doi.org/10.1080/0144929X.2017.1381166

Hannak, A., Sapiezynski, P., Kakhki, A. M., Krishnamurthy, B., Lazer, D., Mislove, A., & Wilson, C. (2013). Measuring personalization of web search. *Proceedings of the 22nd international conference on World Wide Web*, 527–538. https://doi.org/10.1145/2488388.248843

Kiss, Á., & Simonovits, G. (2014). Identifying the bandwagon effect in two-round elections. *Public Choice*, *160*(3), 327–344. https://doi.org/10.1007/s11127-013-0146-y

Kliman-Silver, C., Hannak, A., Lazer, D., Wilson, C., & Mislove, A. (2015). Location, location, location: The impact of geolocation on web search personalization. *Proceedings of the 2015 Internet Measurement Conference*, 121–127. https://doi.org/10.1145/2815675.2815714

Knobloch-Westerwick, S., Johnson, B. K., & Westerwick, A. (2015). Confirmation bias in online searches: Impacts of selective exposure before an election on political attitude strength and shifts. *Journal of Computer-Mediated Communication*, *20*(2), 171–187. https://doi.org/10.1111/jcc4.12105

Knobloch-Westerwick, S., Mothes, C., Johnson, B. K., Westerwick, A., & Donsbach, W. (2015). Political online information searching in Germany and the United States: Confirmation bias, source credibility, and attitude impacts. *Journal of Communication*, *65*(3), 489–511. https://doi.org/10.1111/jcom.12154

Landis, J. R., & Koch, G. G. (1977). The measurement of observer agreement for categorical data. *Biometrics*, *33*(1), 159–174. https://doi.org/10.2307/2529310

Lee, B., Kim, J., & Scheufele, D. A. (2016). Agenda setting in the Internet age: The reciprocity between online searches and issue salience. *International Journal of Public Opinion Research*, *28*(3), 440–455. https://doi.org/10.1093/ijpor/edv026

Makhortykh, M., Urman, A., & Ulloa, R. (2020). How search engines disseminate information about COVID-19 and why they should do better. *Harvard Kennedy School Misinformation Review*, *1*(3). https://doi.org/10.37016/mr-2020-017

Makhortykh, M., Urman, A., & Ulloa, R. (2022). Memory, counter-memory and denialism: How search engines circulate information about the Holodomor-related memory wars. *Memory Studies*, *15*(6), 1330–1345. https://doi.org/10.1177/17506980221133732

Menchen-Trevino, E., Struett, T., Weeks, B. E., & Wojcieszak, M. (2023). Searching for politics: Using real-world web search behavior and surveys to see political information searching in context. *The Information Society*, *39*(2), 98–111. https://doi.org/10.1080/01972243.2022.2152915

Pan, B., Hembrooke, H., Joachims, T., Lorigo, L., Gay, G., & Granka, L. (2007). In Google we trust: Users' decisions on rank, position, and relevance. *Journal of Computer-Mediated Communication*, *12*(3), 801–823. https://doi.org/10.1111/j.1083-6101.2007.00351.x

Pariser, E. (2011). *The filter bubble: What the Internet is hiding from you*. Penguin UK.

Pew Research Center. (2018). *Public has mixed expectations for new tax law*. https://www.pewresearch.org/politics/2018/01/24/public-has-mixed-expectations-for-new-tax-law/

Robertson, R. E., Green, J., Ruck, D. J., Ognyanova, K., Wilson, C., & Lazer, D. (2023). Users choose to engage with more partisan news than they are exposed to on Google Search. *Nature*, *618*(7964), Article 7964. https://doi.org/10.1038/s41586-023-06078-5

Robertson, R. E., Lazer, D., & Wilson, C. (2018). Auditing the personalization and composition of politically-related search engine results pages. *Proceedings of the 2018 World Wide Web Conference*, 955–965. https://doi.org/10.1145/3178876.3186143

Rohrbach, T., Makhortykh, M., & Sydorova, M. (2024). *Finding the white male: The prevalence and consequences of algorithmic gender and race bias in political Google searches* (arXiv:2405.00335). arXiv. https://doi.org/10.48550/arXiv.2405.00335

Röttger, P., Hofmann, V., Pyatkin, V., Hinck, M., Kirk, H. R., Schütze, H., & Hovy, D. (2024). *Political compass or spinning arrow? Towards more meaningful evaluations for values and opinions in large language models* (arXiv:2402.16786). arXiv. https://doi.org/10.48550/arXiv.2402.16786

Rubin, D. B. (1987). *Multiple imputation for nonresponse in surveys* (1st ed.). Wiley. https://doi.org/10.1002/9780470316696

Schmitt-Beck, R. (2015). Bandwagon effect. In *The International Encyclopedia of Political Communication* (pp. 1–5). John Wiley & Sons, Ltd. https://doi.org/10.1002/9781118541555.wbiepc015

Schmuck, D., Tribastone, M., Matthes, J., Marquart, F., & Bergel, E. M. (2020). Avoiding the other side? *Journal of Media Psychology*, *32*(3), 158–164. https://doi.org/10.1027/1864-1105/a000265

Slechten, L., Courtois, C., Coenen, L., & Zaman, B. (2022). Adapting the selective exposure perspective to algorithmically governed platforms: The case of Google Search. *Communication Research*, *49*(8), 1039–1065. https://doi.org/10.1177/00936502211012154

Spink, A., Jansen, B. J., Blakely, C., & Koshman, S. (2006). A study of results overlap and uniqueness among major Web search engines. *Information Processing & Management*, *42*(5), 1379–1391. https://doi.org/10.1016/j.ipm.2005.11.001

The Federal Council. (2024a, März 3). *Initiative for a 13th OASI pension payment*. https://www.admin.ch/gov/en/start/dokumentation/abstimmungen/20240303/iniziativa-per-una-13esima-mensilita-avs.html

The Federal Council. (2024b, März 3). *Pensions Initiative (retirement age)*. https://www.admin.ch/gov/en/start/dokumentation/abstimmungen/20240303/iniziativa-sulle-pensioni.html

Trielli, D., & Diakopoulos, N. (2022). Partisan search behavior and Google results in the 2018 U.S. midterm elections. *Information, Communication & Society*, *25*(1), 145–161. https://doi.org/10.1080/1369118X.2020.1764605

Urman, A., Makhortykh, M., & Ulloa, R. (2022). Auditing the representation of migrants in image web search results. *Humanities and Social Sciences Communications*, *9*(1), 1–16. https://doi.org/10.1057/s41599-022-01144-1

van Buuren, S. (2018). *Flexible imputation of missing data* (2nd ed.). Chapman and Hall/CRC. https://doi.org/10.1201/9780429492259

van der Eijk, C., van der Brug, W., Kroh, M., & Franklin, M. (2006). Rethinking the dependent variable in voting behavior: On the measurement and analysis of electoral utilities. *Electoral Studies*, *25*(3), 424–447. https://doi.org/10.1016/j.electstud.2005.06.012

van Ginkel, J. R., Linting, M., Rippe, R. C. A., & van der Voort, A. (2020). Rebutting existing misconceptions about multiple imputation as a method for handling missing data. *Journal of Personality Assessment*, *102*(3), 297–308. https://doi.org/10.1080/00223891.2018.1530680

van Hoof, M., Meppelink, C. S., Moeller, J., & Trilling, D. (2024). Searching differently? How political attitudes impact search queries about political issues. *New Media & Society*, *26*(7), 3728-3750. https://doi.org/10.1177/14614448221104405

von Hippel, P. T. (2020). How many imputations do you need? A two-stage calculation using a quadratic rule. *Sociological Methods & Research*, *49*(3), 699–718. https://doi.org/10.1177/0049124117747303

Westerwick, A., Johnson, B. K., & Knobloch-Westerwick, S. (2017). Confirmation biases in selective exposure to political online information: Source bias vs. content bias. *Communication Monographs*, *84*(3), 343–364. https://doi.org/10.1080/03637751.2016.1272761

Zumofen, G. (2023). What drives the selection of political information on Google? Tension between ideal democracy and the influence of ranking. *Swiss Political Science Review*, *29*(1), 120–138. https://doi.org/10.1111/spsr.12545